\documentclass[twocolumn,superscriptaddress,floatfix,preprintnumbers,amssymb ,amsmath]{revtex4}
\usepackage{graphicx}
\usepackage{dcolumn}
\usepackage{bm}
\usepackage[latin1]{inputenc}
\usepackage[mathscr]{eucal}
\usepackage{epsfig}

\begin{document}

\title{Electron-phonon coupling and longitudinal mechanical-mode cooling in a metallic nanowire}

\author{F.W.J. Hekking}
\affiliation{Laboratoire de Physique et Mod\'elisation des Milieux
Condens\'es, \\ C.N.R.S. and Universit\'e Joseph Fourier, B.P. 166,
38042 Grenoble Cedex 9, France}
\author{A.O. Niskanen}
\affiliation{VTT Technical Research Centre of Finland, Sensors, PO
BOX 1000, 02044 VTT, Finland}
\author{J.P. Pekola}
\affiliation{Low Temperature Laboratory, Helsinki University of Technology,
PO BOX 3500, 02015 TKK, Finland}

\begin{abstract}
We investigate electron-phonon coupling in a narrow suspended
metallic wire, in which the phonon modes are restricted to one
dimension but the electrons behave three-dimensionally. Explicit
theoretical results related to the known bulk properties are
derived. We find out that longitudinal vibration modes can be cooled
by electronic tunnel refrigeration far below the bath temperature
provided the mechanical quality factors of the modes are
sufficiently high. The obtained results apply to feasible
experimental configurations.
\end{abstract}





\date{\today}

\maketitle

Electron-phonon coupling in metals, albeit extensively studied over
several decades \cite{gantmakher74}, is of utmost interest and
importance in view of present day developments in
nanoelectromechanics \cite{cleland03,schwab05} and in electronic
cooling and sensing on nanoscale \cite{giazotto06,saira07}. A number
of questions arise when the dimensionality of the phonons is reduced
from the conventional bulk three-dimensional case
\cite{qu05,yu95,kuhn06}. Recent experimental observations of
metallic wires on thin dielectric membranes support the fact that
reduction of phonon dimensionality leads to weaker temperature
dependence of the heat flux between electrons and phonons
\cite{karvonen07}. Very little is known about truly one-dimensional
wires, where transverse dimensions are far smaller than the thermal
wavelength of the phonons, although this regime is readily available
experimentally at sub-kelvin temperatures in wires whose diameter is
of the order of 100 nm or less. Recently though, substantial
overheating was conjectured to be the origin of excess low-frequency
charge noise in a suspended single-electron transistor in this
particular one-dimensional geometry \cite{li07}. In this Letter we
derive an explicit result for electron-phonon heat flux in a
metallic wire in which electrons behave three-dimensionally but
phonons are confined to one dimension, and relate this result to the
standard bulk result for the corresponding metal. We present a
scenario of tunnel coupling the metal electrons in a wire to a
superconductor on bulk, whereby cooling of wire electrons can be
realized. We demonstrate that the few available mechanical modes,
i.e., discrete longitudinal phonons, can be cooled significantly by
their coupling to the cold electrons in the wire. This occurs
provided the mechanical modes are not too strongly coupled to the
thermal bath, meaning that the mechanical $Q$ value of the mode is
sufficiently high. Recently, indirect experimental evidence of
electronic cooling of phonons in a bulk system was put forward in
Ref. \cite{rajauria07}.

\begin{figure}
\begin{center}
\includegraphics[width=0.5
\textwidth]{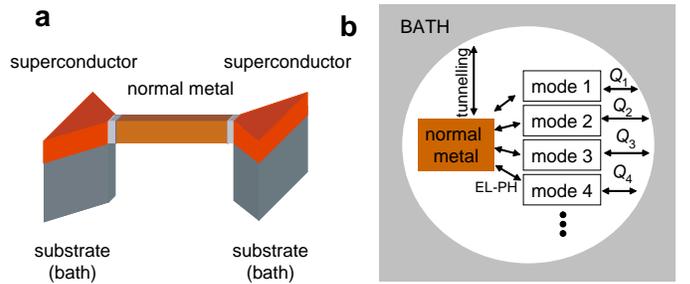} \caption{The system under study. In (a) we
show the suspended wire whose transverse dimensions are supposed to
be smaller than the thermal wavelength of the phonons, $\lambda_{\rm
thermal}$. In this particular example the metal wire is connected to
the bulk superconducting reservoirs via tunnel barriers to form a
tunnel junction refrigerator. In (b) we show the relevant thermal
model of the system.}\label{Fig1}
\end{center}
\end{figure}
To obtain results for the electron-phonon heat flux in a
one-dimensional metallic wire (see Fig. \ref{Fig1} for the geometry
and thermal model), we follow the standard procedure from the
existing literature normally applied to the case of either bulk
three-dimensional phonons \cite{gantmakher74,wellstood94}, or to the
case where phonons are restricted to a semi-infinite bulk
\cite{qu05}. The net heat flux from electrons into a discrete phonon
mode $\mu$ at wave vector ${\bf q}$ is given by
\begin{equation} \label{flux}
\dot{Q}_{e\rightarrow \mu}({\bf q})= 2\sum_{\bf k}\hbar \omega_\mu
[\Gamma^{\rm e}_\mu({\bf k}\rightarrow {\bf k - q})- \Gamma^{\rm
a}_\mu({\bf k}\rightarrow {\bf k + q})],
\end{equation}
where phonon emission (e) and absorption (a) rates by the electrons
with wave vector ${\bf k}$ are obtained via the golden rule as
\begin{eqnarray} \label{em}
&& \Gamma^{\rm e}_\mu({\bf k}\rightarrow {\bf k - q})=
\frac{2\pi}{\hbar} |g_{\mu,\bf q}|^2[n(\frac{\hbar \omega_\mu}{k_B
T_\mu})+1]\nonumber \\
&& \times f(E_{\bf k})[1-f(E_{\bf k - q})]\delta(E_{\bf k}-E_{\bf
k - q}-\hbar \omega_\mu),
\end{eqnarray}
and
\begin{eqnarray} \label{ab}
&& \Gamma^{\rm a}_\mu({\bf k}\rightarrow {\bf k + q})=
\frac{2\pi}{\hbar} |g_{\mu,\bf q}|^2n(\frac{\hbar \omega_\mu}{k_B
T_\mu})\nonumber \\
&& \times f(E_{\bf k})[1-f(E_{\bf k + q})]\delta(E_{\bf k}-E_{\bf
k + q}+\hbar \omega_\mu).
\end{eqnarray}
Here $g_{\mu,\bf q}$ and $n(\frac{\hbar \omega_\mu}{k_B
T_\mu})=[\exp(\frac{\hbar\omega_\mu}{k_B T_\mu})-1]^{-1}$ are the
electron-phonon coupling constant and the Bose distribution,
respectively, of the phonon mode $\mu$ at angular frequency
$\omega_\mu$ and at temperature $T_\mu$, and
$f(E)=[\exp(\frac{E}{k_B T_e})+1]^{-1}$ is the Fermi distribution
of the electrons at temperature $T_{e}$.

We next evaluate $g_{\mu,\bf q}$ using the standard results of
deformation potential of the collective lattice vibrations
\cite{cleland03}. Let ${\bf w}_\mu({\bf r})$ be the displacement
vector for mode $\mu$ normalized in the volume of the wire,
$\mathcal{V}$, such that
$\int d^3{\bf r}\,\, {\bf w}_\mu({\bf r}){\bf \cdot} {\bf
w}_{\mu'}^*({\bf r}) = \delta_{\mu \mu'}$.
Then $g_{\mu,\bf q}$ can be obtained from the divergence of ${\bf
w}_\mu({\bf r})$
\begin{equation} \label{cpl}
g_{\mu,\bf q}= \frac{2}{3}E_F\sqrt{\frac{\hbar}{2\rho
\omega_\mu}}\int d^3{\bf r}\,\, \Psi_{\bf k-q}({\bf r}) \Psi_{\bf k}^\dagger({\bf r})
\,\, \nabla \cdot{{\bf w}_\mu({\rm r})},
\end{equation}
where $\Psi_{\bf k}({\bf r}) $ are the electronic wave functions.
Here $E_F$ is the Fermi energy of the electrons and $\rho$ is the
mass density of the wire.

The thermal wavelength of phonons, $\lambda_{\rm thermal}=\frac{h
c_\ell}{k_B T}$ at temperature $T$ and mode velocity $c_\ell$ is
typically of order 1 $\mu$m at $T=100$ mK. In a wire whose length $L
> \lambda_{\rm thermal}$ and with transverse dimensions $\ll \lambda_{\rm thermal}$, only modes
with ${\bf q} = (0,0,q)$ directed along the wire ($z$-axis) appear
relevant, since the ones with perpendicular ${\bf q}$ are too high
in energy. There are basically four types of vibrations:
longitudinal, flexural (two, with $x$ and $y$ polarizations),
torsional and shear modes \cite{rego98}. The last one has a gap and
is therefore not excited at low temperatures. Of the remaining ones
the torsional modes have no divergence, and essentially only the
longitudinal modes couple to electrons in the long wavelength limit.
Experimentally this seems to be the case in carbon nanotubes
\cite{sapmaz06}.

We consider longitudinal modes with specific boundary conditions:
the wire (or the three-dimensional body) is assumed to be clamped
at the ends. As we will detail below, this corresponds to a
feasible realization.
Let the wire extend from $z=0$ to $z=L$. Then, the normalized
longitudinal eigenmodes of the beam are given by
\begin{equation}
{\bf w}_\mu({\bf r})  =\sqrt{\frac{2}{\mathcal{V}}} \sin(\mu \pi
z/L)\hat{\bf z} \mbox{, } \mu = 1,2,3, \ldots
\end{equation}
They are characterized by the linear dispersion relation $\omega_\mu
= c_\ell \mu \pi/L$, where $c_\ell=\sqrt{E/\rho}$ is the
longitudinal sound velocity ($E$ is Young's modulus). Assuming zero
electronic boundary conditions along with equal electronic and
phononic volumes we obtain again in the long wavelength limit
\begin{equation} \label{cplfbc}
|g_{\mu,\bf q}|^2= \frac{1}{9} \frac{\hbar E_F^2 q^2}{\rho
\mathcal{V}\omega_\mu} \delta_{q,q_\mu}\equiv \mathcal{M}_\ell^2 q
\delta_{q,q_\mu}, \,\,q_\mu=\mu\pi/L,
\end{equation}
where $\mathcal{M}_\ell^2 \equiv \frac{1}{9} \frac{\hbar
E_F^2}{\rho \mathcal{V}c_\ell}$. The momentum $q$ transferred
between the electron and the vibrational modes of a clamped beam
takes discrete values $q_\mu$ only and is by convention positive.

We perform next the integration over electron energies in
Eqs.~\eqref{em} and \eqref{ab} and insert the results in
Eq.~\eqref{flux} obtaining
\begin{equation} \label{fluxmode}
\dot{Q}_{e\rightarrow \mu}(q_\mu)= \frac{2\pi \mathcal{M}_\ell^2
c_\ell^2 m N(E_F)}{\hbar k_F} q_\mu^2[n(\frac{\hbar c_\ell
q_\mu}{k_BT_e})-n(\frac{\hbar c_\ell q_\mu}{k_BT_{\mu}})].
\end{equation}
Here, $m$ is the electron mass, $k_F$ the Fermi wave vector, and
$N(E_F)$ the electronic density of states at the Fermi energy.
Three-dimensional distribution of electrons was assumed here, since
we discuss only the case of ordinary metals, where $k_F^{-1} \ll 1$
nm. Using the definition of $\mathcal{M}_\ell^2$ above, and $N(E_F)=
\frac{mk_F\mathcal{V}}{\pi^2 \hbar^2}$ and $E_F
=\frac{\hbar^2k_F^2}{2m}$ of the free electron gas, the prefactor in
Eq. \eqref{fluxmode} can also be written in the form
$\frac{2\pi \mathcal{M}_\ell^2 c_\ell^2 m N(E_F)}{\hbar k_F}=
\frac{1}{18\pi}\frac{\hbar^2 k_F^4 c_\ell}{\rho}$.
The total heat flux between electrons and phonons can then be
obtained as a sum over all modes:
\begin{equation} \label{fluxint}
\dot{Q}_{e\rightarrow p}= \sum_{\mu} \dot{Q}_{e\rightarrow
\mu}(q_\mu).
\end{equation}

We obtain the continuum result for a long $L \gg \lambda_{\rm
thermal}$ 1D wire by assuming a uniform density of modes with all
of them at the same temperature $T_\mu = T_p$. We then replace the
sum by an integral, $\sum_q \rightarrow \frac{L}{\pi}\int_0^\infty
dq$. After a straightforward integration we obtain
\begin{equation} \label{q1d}
\dot{Q}_{e\rightarrow p} = \Sigma_{\rm 1D} L (T_e^3 -T_p^3).
\end{equation}
Here, $\Sigma_{\rm 1D}$ is given by
\begin{equation} \label{sigma1d}
\Sigma_{\rm 1D} = \frac{\zeta(3)}{18\pi^2}\frac{k_F^4k_B^3}{\hbar
c_\ell^2\rho}.
\end{equation}
It is instructive to compare this result to the celebrated result
for longitudinal phonons in three dimensions (see, e.g.,
\cite{wellstood94} and references therein)
\begin{equation} \label{fluxbulk}
\dot{Q}_{e\rightarrow p} = \Sigma \mathcal{V} (T_e^5 -T_p^5).
\end{equation}
Here, the material specific prefactor $\Sigma$ is given by:
\begin{equation} \label{sigma3d}
\Sigma = \frac{\zeta(5)}{3\pi^3}\frac{k_F^4k_B^5}{\hbar^3
c_\ell^4\rho}.
\end{equation}
We conclude that $\Sigma_{\rm 1D}$ is related to the known
$\Sigma$ of the bulk by
\begin{equation} \label{s1s3}
\Sigma_{\rm 1D}
=\frac{\pi}{6}\frac{\zeta(3)}{\zeta(5)}(\frac{\hbar
c_\ell}{k_B})^2 \Sigma.
\end{equation}
Note that Eq.~\eqref{q1d} with the relation \eqref{s1s3} between
$\Sigma_{1D}$ and $\Sigma$ are quite general, and do not depend on
the choice of free electron gas parameters that lead to
Eqs.~\eqref{sigma3d} and \eqref{sigma1d}. Equation \eqref{q1d}
with the help of \eqref{s1s3} and the experimentally determined
$\Sigma$ can then be used to assess electron-phonon coupling in
one-dimensional wires. Equation \eqref{sigma3d} predicts the
behaviour of real metals rather well: the overall magnitude of
$\Sigma$ from \eqref{sigma3d} with parameters of usual metals is
of order $\Sigma \sim 10^8$ WK$^{-5}$m$^{-3}$, whereas measured
values are typically around $10^9$ WK$^{-5}$m$^{-3}$. The
deviation may be partly ascribed to the complicated structure of
the Fermi surface in real metals \cite{qu05}.

Equations \eqref{fluxbulk} and \eqref{q1d} predict correctly the
cross-over between three-dimensional and one-dimensional behaviour.
To see this, let us look at the linearized heat conductance for a
small temperature difference $\Delta T \equiv T_e - T_p$ between
electrons and phonons, such that $\dot{Q}_{e\rightarrow p} \simeq
G_{ep}\Delta T$. From Eq. \eqref{fluxbulk}, we obtain $G_{ep}^{\rm
3D} = 5\Sigma\mathcal{V}T^4$, where we denote by $T$ the (almost)
common temperature of the two subsystems. Similarly from Eq.
\eqref{q1d} we obtain $G_{ep}^{\rm 1D} = 3\Sigma_{\rm 1D}LT^2$. Now
let us consider a wire whose square cross-section is $w\times w$.
The cross-over between 3D and 1D behaviour is expected to occur when
the first longitudinal modes get occupied thermally within the
cross-section, i.e., when $\hbar c_\ell /w \sim k_B T$. Making use
of the relation \eqref{s1s3}, and $\mathcal{V}\equiv Lw^2$, we then
see that with the above condition the expressions of $G_{ep}^{\rm
3D}$ and $G_{ep}^{\rm 1D}$ become identical in form, apart from
numerical prefactors.

Next we demonstrate that variation of electron temperature in the
wire leads to variation of the temperature of its vibrational
modes. In particular, electron mediated mechanical mode cooling
becomes possible. If we assume a highly underdamped mechanical
mode whose quality factor $Q_\mu\gg1$, we can obtain the heat flux
from the thermal bath into the mode $\mu$ in a classical picture
as
\begin{equation} \label{classQ}
\dot{Q}_{{\rm bath} \rightarrow \mu} = \frac{k_B
\omega_n}{Q_\mu}(T_{\rm bath}-T_{\mu}).
\end{equation}
This result can be inferred as a solution of the Fokker-Planck
equation of Brownian motion in the harmonic potential or by direct
solution of Langevin equation \cite{reif65}. We have assumed that
the mode temperature is given by the equipartition principle via
$k_B T_\mu = k \langle x^2 \rangle$ for the position $x$ of the
Brownian particle with spring constant $k$. Equation
\eqref{classQ} is the high temperature limit of the quantum
expression of heat flux
\begin{equation} \label{quantQ}
\dot{Q}_{{\rm bath} \rightarrow \mu} = \frac{\hbar
c_\ell^2}{Q_\mu} q_\mu^2[n(\frac{\hbar c_\ell q_\mu}{k_B T_{{\rm
bath}}})-n(\frac{\hbar c_\ell q_\mu}{k_B T_{\mu}})],
\end{equation}
which is identical in form with Eq. \eqref{fluxmode}. We have
again identified $\omega_\mu = c_\ell q_\mu$. One then finds a
steady-state temperature of the mode $\mu$ by solving the balance
equation (see Fig. \ref{Fig1}),
\begin{equation} \label{fluxbathf}
\dot{Q}_{{\rm bath} \rightarrow \mu}+\dot{Q}_{e\rightarrow \mu}=0.
\end{equation}
There are some interesting limits: if $\frac{\hbar c_\ell^2}{Q_\mu}
\ll \frac{1}{18\pi}\frac{\hbar^2 k_F^4 c_\ell}{\rho}$, electrons
cool efficiently and the mode temperature follows $T_e$, whereas in
the opposite limit the mode temperature stays at $T_{\rm bath}$.
Eliminating $k_F$ in favour of experimentally determined $\Sigma$,
we find that the temperature of the mechanical mode follows that of
the electrons if
$Q_\mu \gg \frac{12 \zeta(5)}{\pi^2} \frac{k_B^5}{\hbar^4 c_\ell^3
\Sigma}$.
With parameters of ordinary metals this leads to the condition
$Q_\mu \gg 100$.

We conclude the formal part by obtaining a useful relation yielding
the heat flux between electrons and the bath with the help of their
respective temperatures, using Eqs. \eqref{fluxmode},
\eqref{fluxint}, \eqref{quantQ} and \eqref{fluxbathf}, and assuming
that all the relevant modes have the same quality factor $Q$:
\begin{eqnarray}
&& \dot{Q}_{e \rightarrow {\rm bath}} =
\frac{\frac{\pi^2}{12\zeta(5)}(\frac{\hbar c_\ell}{k_B})^5\Sigma}{1+
\frac{\pi^2}{12\zeta(5)}\frac{\hbar^4
c_\ell^3}{k_B^5}Q\Sigma}\nonumber \\
&& \times \sum_\mu q_\mu^2 [n(\frac{\hbar c_\ell q_\mu}{k_B
T_{e}})-n(\frac{\hbar c_\ell q_\mu}{k_B T_{\rm bath}})].
\end{eqnarray}
An expression of type \eqref{q1d} can be obtained in the continuum
limit again, but here the factor $\Sigma_{\rm 1D}$ must be replaced
by $[1+ \frac{\pi^2}{12\zeta(5)}\frac{\hbar^4
c_\ell^3}{k_B^5}Q\Sigma]^{-1}\Sigma_{\rm 1D}$.

\begin{figure}
\begin{center}
\includegraphics[width=0.5
\textwidth]{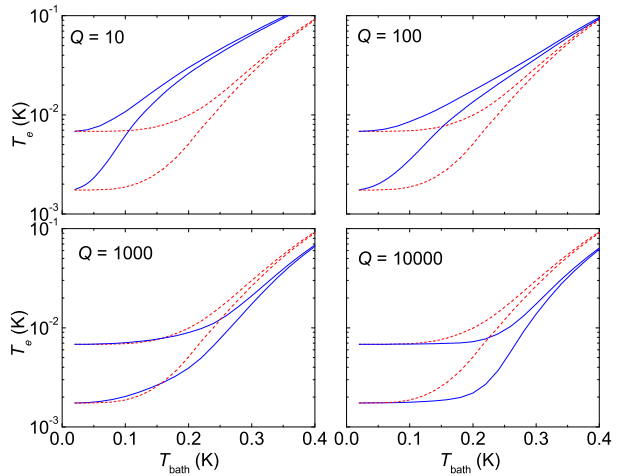} \caption{Minimum electron temperature in a
metal wire as a function of bath temperature, cooled by SINIS
tunnelling for various values of $Q$. The solid blue lines are from
the one-dimensional model, and the red dashed lines from the
three-dimensional model. The parameters we used correspond to Al as
a superconductor and Cu as the metal wire of $L=1$ $\mu$m length;
its width and thickness are assumed to be 30 nm both,
$\Sigma=2\cdot10^9$ WK$^{-5}$m$^{-3}$, $E=130$ GPa and $\rho=8920$
kgm$^{-3}$. The tunnel resistances of the two NIS junctions are $R_T
= 10$ k$\Omega$ both, and we assume that the non-ideality parameter
of the junctions has a value $\gamma = 1\cdot 10^{-4}$ (curves lying
higher) or $\gamma = 1\cdot 10^{-5}$ (lower).}\label{Fig2}
\end{center}
\end{figure}

\begin{figure}
\begin{center}
\includegraphics[width=0.5
\textwidth]{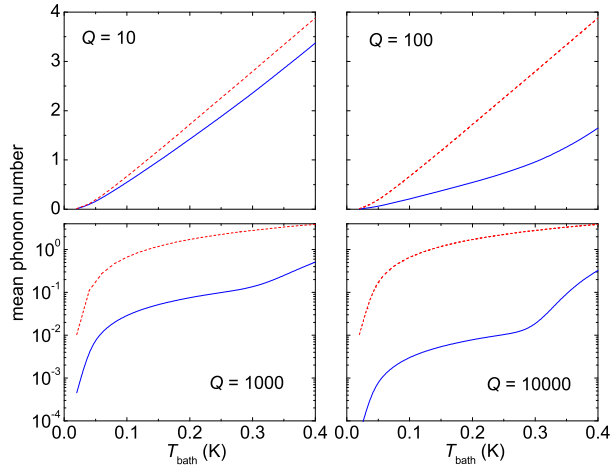} \caption{Population of the lowest vibrational
mode in a wire against bath temperature according to the
one-dimensional model. The solid (blue) lines show the population
when the electron cooling is applied, whereas the dashed (red) lines
are the corresponding non-cooled populations. The parameters are the
same as in Fig. \ref{Fig2}. The results for the two values of
$\gamma$ are almost indistinguishable, and we have plotted only
those for $\gamma = 1\cdot 10^{-4}$.}\label{Fig3}
\end{center}
\end{figure}

We next apply the results above to determine the performance and
mechanical mode cooling \cite{naik06,schliesser06,poggio07,brown07}
in a suspended electron refrigerator. Note that overheating of a
suspended wire, or a single-electron transistor \cite{li07}, can be
analyzed similarly as our example of cooling below: heat currents
and temperature drops are simply inverted. In a hybrid tunnel
junction configuration (SINIS), with a metal island (N) and
superconducting leads (S), the electron system in N can be cooled
far below the bath temperature by applying a bias $eV$ of the order
of the superconducting gap $\Delta$ over each tunnel junction (I)
between S and N. This SINIS refrigeration technique based on energy
filtering of the tunnelling electrons due to the gap in the
superconductor has been applied extensively over the past decade,
for a review see Ref. \cite{giazotto06}, but not yet in suspended
wires to the best of our knowledge. Here we propose its use in
connection with the one-dimensional phonon system. It is possible to
cool not only the electrons in the wire but also the vibrational
modes in it by coupling them to the cold electrons. Figure
\ref{Fig2} shows numerically calculated results for the minimum
electron temperature reached as a function of the bath temperature:
at the optimum bias voltage of the junctions heat is removed from
the wire at a rate $\dot{Q} \sim \Delta^2/(e^2R_T)(T_e/T_C)^{3/2}$.
In steady state this heat flux is balanced by the heat flux from the
phonon modes. We assume that all the relevant modes have the same
quality factor $Q_n \equiv Q$. The collection of results in Fig.
\ref{Fig2} shows that if $Q$ is large, strong suppression of
electron temperature can be achieved. The saturation of the
temperature with low $T_{\rm bath}$ is caused by the ohmic heating
in the refrigerating junctions with leakage parameter $\gamma$,
which has been chosen to correspond to typical experimental
conditions: $\gamma$ equals the low temperature zero bias
conductance of a junction normalized by the value of conductance at
large voltages, and it can be conveniently included in the
(normalized) density of quasiparticle states of the superconductor
at energy $E$ as $n_S(E)=|{\rm Re}(\frac{E+i\gamma
\Delta}{\sqrt{(E+i\gamma \Delta)^2 -\Delta^2}})|$ \cite{giazotto06}.
The cooling effect of the suspended structure differs from that of
the result of the three-dimensional model; specifically the results
of the one-dimensional model, valid when $w \lesssim \lambda_{\rm
thermal}$, do not depend on the transverse dimensions of the wire,
whereas the results of the three-dimensional model are determined by
these dimensions as well via the dependence on volume in Eq.
\eqref{fluxbulk}. Also the vibrational modes involved are cooled:
this is demonstrated in Fig. \ref{Fig3}, where we plot the
population of the lowest mode, $n=1$, under the same conditions as
in Fig. \ref{Fig2}. The corresponding mode occupations in the
absence of electron cooling are shown for reference.  The magnitude
of the mode cooling is determined by the interplay of the cooling
power, electron-phonon coupling, and the coupling to the bath,
determined by $Q$. From our example it seems obvious that
electron-mediated cooling of the vibrational modes into the quantum
limit is a feasible option, manifested by the very low mode
populations, in particular when $Q$ is large.

In summary, we derived the basic relations governing electron-phonon
heat transport in narrow metal wires, where the electron
distribution is three-dimensional and the phonon distribution is
confined to one dimension. In this realistic scenario describing
suspended wires made of ordinary metals, we find that the heat
currents differ drastically from those in bulk systems. In
particular, we demonstrated that the vibrational modes of the wire
can be cooled significantly by electron refrigeration, provided the
mechanical $Q$'s of the modes are sufficiently high.

We thank the NanoSciERA project "NanoFridge" of the EU and the
Academy of Finland for financial support.

\end{document}